\documentclass[%
reprint,
superscriptaddress,
 amsmath,amssymb,
 aps,
 prl,
]{revtex4-2}

\usepackage{graphicx}
\usepackage{dcolumn}
\usepackage{bm}

\usepackage[cp1250]{inputenc}
\usepackage{ulem,color}

\newcommand{\del}[1]{}

\newcommand{\cc}[1]{}

\begin{document}

\preprint{APS/123-QED}

\title{Plasmonic lightning-rod effect}

\author{Vlastimil Křápek}
\affiliation{Brno University of Technology, Central European Institute of Technology, Purkyňova 123, 612 00 Brno, Czechia}
\affiliation{Brno University of Technology, Institute of Physical Engineering, Technická 2, 616 69 Brno, Czechia}
\email{krapek@vutbr.cz}
\author{Rostislav Řepa}
\author{Michael Foltýn}
\affiliation{Brno University of Technology, Institute of Physical Engineering, Technická 2, 616 69 Brno, Czechia}
\author{Tomáš Šikola}
\author{Michal Horák}
\affiliation{Brno University of Technology, Central European Institute of Technology, Purkyňova 123, 612 00 Brno, Czechia}
\affiliation{Brno University of Technology, Institute of Physical Engineering, Technická 2, 616 69 Brno, Czechia}
\email{michal.horak2@ceitec.vutbr.cz}

\date{\today}

\begin{abstract}
The plasmonic lightning-rod effect refers to the formation of a strong electric near field of localized surface plasmons at the sharp features of plasmonic antennas. While this effect is intuitively utilized in the design and optimization of plasmonic antennas, the relation between the magnitude of the electric field and the local curvature of the plasmonic antenna has not been yet rigorously established. Here, we provide such a study. We design sets of plasmonic antennas that allow to isolate the role of the local curvature from other effects influencing the field.
The near electric field is inspected by electron energy loss spectroscopy and electrodynamic simulations. We demonstrate the existence of the plasmonic lightning-rod effect and establish its quantitative description, showing that its strength is comparable to the electrostatic lightning-rod effect. We also provide a simple phenomenological formula for the spatial dependence of the field. Finally, we introduce the effective radius of curvature related to the spatial distribution of induced charge in plasmonic antennas, significantly smaller than their geometrical radius. \cc{druha iterace}
\end{abstract}


\maketitle

{\it Introduction} Plasmonic antennas~\cite{Novotny2011} (PA) represent a class of open photonic cavities formed by metallic nanostructures that support localized surface plasmons (LSP). The excitation of LSP is related to the formation of a spatially confined and enhanced electromagnetic field, which gives rise to enhanced light-matter interaction. This effect is utilized in a wide range of applications including plasmon-enhanced spectroscopy (photoluminescence spectroscopy~\cite{Kinkhabwala2009,Pfeiffer2010}, Raman spectroscopy~\cite{Wang2020,Han2022}, electron paramagnetic resonance spectroscopy~\cite{https://doi.org/10.1002/smtd.202100376}), energy harvesting~\cite{BORISKINA2013375}, strong light-matter coupling~\cite{doi:10.1021/nl104352j,Bitton2020}, sensing~\cite{annurev:/content/journals/10.1146/annurev.physchem.58.032806.104607,Mejia-Salazar2018}, and medicine~\cite{https://doi.org/10.1002/lpor.201200003,Shrivastav2021}.

The design and optimization of PAs exploit ingenious concepts including transformation optics~\cite{doi:10.1021/nl303377g,doi:10.1021/acs.jpcc.2c04828}, plasmonic circuit models~\cite{PhysRevLett.95.095504,Zhu:14,Benz:15,Hughes2016}, Babinet's principle~\cite{PhysRevB.76.033407,doi:10.1021/nl402269h,Horak2019,10.1063/5.0065724}, hybridization models~\cite{doi:10.1126/science.1089171,krapek_independent}, as well as material aspects~\cite{Khurgin2012}. One of the widely utilized concepts is the so-called plasmonic lightning-rod effect (PLRE), which refers to the local enhancement of the electric field at the parts of PA with a high local curvature.

\cc{... kde se pouziva. Dalsi odstavec - intuitivne se pouziva, rigorozne nebyl studovan - tady bude nase reserse. Zatim prvni verze}

Applications of PLRE include among others energy-storing and energy-generating devices~\cite{10.1063/1.5091723}, multibeam photoemission cathode~\cite{ZHANG2017114}, single photon emission for quantum communication~\cite{Fiore_2007}, or picocavities allowing to study single molecule optomechanics~\cite{doi:10.1126/science.aah5243}. Although widely used, PLRE is not rigorously understood despite studies addressing its particular aspects~\cite{Garcia-Etxarri:12,Zhou2013,Urbieta2018}.

Our study fills this gap by addressing and establishing the relation between the local electric field of LSP and the local curvature of PAs. We identify and control additional possible effects contributing to the field magnitude in order to isolate the pure effect of the curvature.

\begin{figure}
\includegraphics[clip,width=\columnwidth]{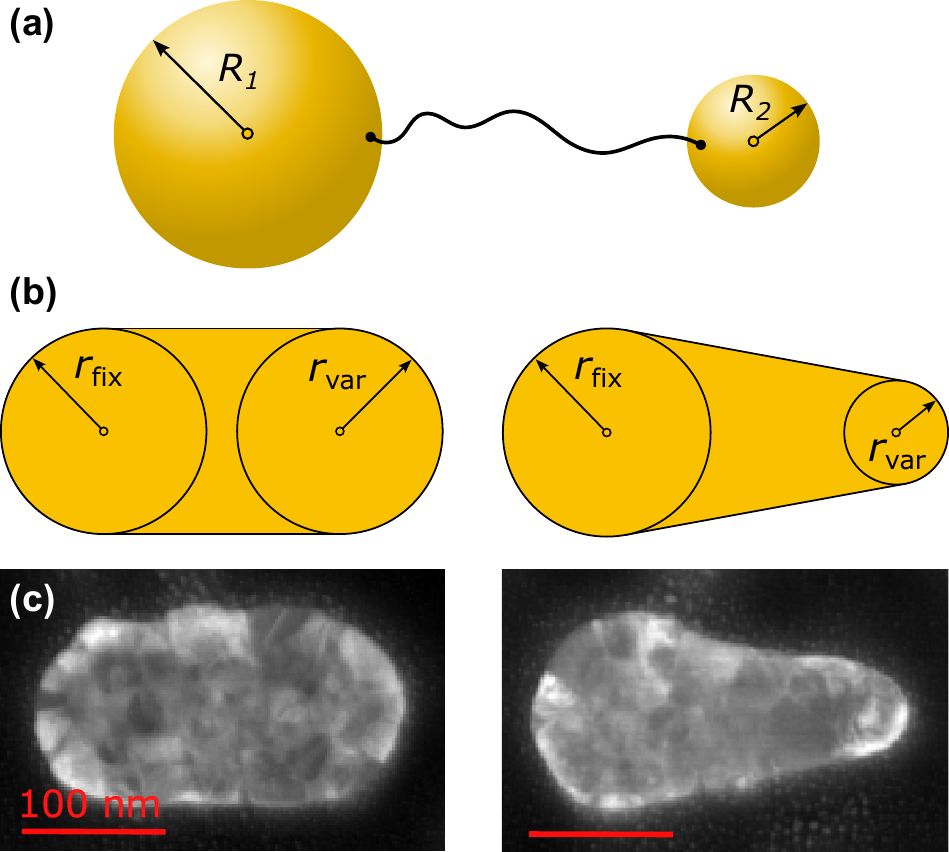}
\caption{\label{fig1}
(a) A scheme of setup demonstrating the electrostatic lightning-rod effect. (b) Design of the plasmonic beetroot -- a platform for studying PLRE. (c) HAADF images of two fabricated PBs with $r_\mathrm{fix}=70~\mathrm{nm}$ and $r_\mathrm{var}=70~\mathrm{nm}$ (left) and 20~nm (right).}
\end{figure}

{\it Design} The lightning rod effect was originally formulated in electrostatics, where it can be illustrated for a system of two distant metallic spheres [Fig.~\ref{fig1}(a)]. For a single homogeneous metallic sphere of the radius $R$ charged with a charge $Q$, the electrostatic potential and electric field on its surface read $\varphi=k_eQ/R$, $E=k_eQ/R^2$, respectively, where $k_e=1/4\pi\varepsilon_0$ is the Coulomb constant. We will now consider two such spheres (labeled with subscripts 1 and 2) with the radii of $R_1$ and $R_2$. The spheres are far enough for the mutual electrostatic interaction to be negligible, and conductively connected with a long thin wire. When the system is charged, the total charge $Q$ redistributes so that the electrostatic potentials of both spheres are equal, $\varphi_1=\varphi_2=\varphi$, yielding $E_i=\varphi/R_i$. The field at the surface of the spheres is inversely proportional to their local radius. For conductors of more complex shapes, this simple quantitative dependence does not hold, but the qualitative trend of larger fields for lower local radii is preserved. 

PLRE differs in two important aspects. First, both propagating and localized surface plasmon polaritons are evanescent waves with naturally confined electric field even for zero curvature of the supporting PA. As an example, we will consider a planar metal-vacuum interface. When such an interface is statically charged with a surface charge density $\sigma$, it produces a homogeneous field in the vacuum with the magnitude $E_\mathrm{stat}=\sigma/\varepsilon_0$. On the other hand, when a SPP is excited on such interface, the related electric field decays exponentially into the vacuum as $E_\mathrm{SPP}(z)=E_0\exp(-\kappa z)$~\cite{hohenester2019nano}, where $z$ denotes the distance from the interface into vacuum and $\kappa=(\omega/c)\cdot[1/\varepsilon_m+1]^{1/2}$ is the decay constant (with $\omega$, $c$, and $\varepsilon_m$ representing the angular frequency of SPP, the speed of light in vacuum, and the dielectric constant of the metal, respectively). 
Second, in electrostatics, the metallic object is charged, while in plasmonics the PA is quasineutral with charge oscillations related to LSP driven by an external excitation. Distinct PAs respond differently to the same excitation, yielding different magnitudes of the induced charge as well as the magnitudes of the induced electric near field. In the following, we denote this as the charge reservoir effect. \cc{Excitation efficiency effect?}

The electric field of PAs is affected by all above-mentioned effects (PLRE, evanescent confinement, charge reservoir), and possibly other effects not yet identified. To isolate the PLRE, we need to either remove or account for the variations of any other effects when changing the local curvature of PAs. To this end, we design the set of planar rod-shaped PAs with a fixed length and two cylindrical terminations [Fig.~\ref{fig1}(b)]. In the following, we will denote these PAs as plasmonic beetroots (PBs). In each set of plasmonic beetroots, one of the terminations has a fixed radius $r_\mathrm{fix}$ and serves as a reference, while the other termination has a variable radius $r_\mathrm{var}$ and is utilized to monitor the PLRE. The fixed length of all PBs in each set shall ensure similar energies of LSP modes and thus exclude variations in the evanescent confinement. Variations in any effects contributing to the electric field other than PLRE shall be reflected in variations of the reference field at the fixed-radius termination.  The electric field recorded at the variable-radius termination normalized by the reference field at the fixed-radius termination shall then reflect solely the effect of the local curvature.

{\it Methods} \cc{je treba overit parametry}
Our methodology combines electromagnetic simulations with experimental electron energy loss spectroscopy (EELS). The simulations are used for a full quantitative vectorial characterization of the electric field while EELS is utilized to verify the simulations and also provides a qualitative insight into the field~\cite{PhysRevLett.100.106804,PhysRevLett.103.106801,C3CS60478K,krapek_independent}.

For the fabrication of PBs, a thin gold film with a thickness of 30 nm is deposited by ion-beam assisted sputtering onto a standard 30-nm-thick TEM membrane (Agar Scientific), and the desired PB shape is achieved by nanofabrication with focused ion beam milling (FEI Helios 660 dual-beam microscope, Ga$^+$ ions at 30 keV). The high-angle annular dark field (HAADF) images of fabricated PBs are shown in Fig.~\ref{fig1}(c). Further structural and chemical parameters of PAs fabricated with this procedure are reported in Refs.~\cite{Horak2018,Kejik:20}.

EELS measurements were carried out in a scanning transmission electron microscope (FEI Titan with GIF Quantum spectrometer; primary beam energy 120 keV, probe current 100~pA, acquisition time 0.5~ms). The full-width at half-maximum of the zero-loss peak was found in the range from 0.10 to 0.12~eV. The raw data were normalized to yield the loss probability, from which the zero-loss peak and the background were subtracted to isolate the contribution of LSP. Individual peaks corresponding to LSP resonances were fitted with Lorentzian spectral profiles.

Theoretical modeling was performed using a software package MNPBEM~\cite{HOHENESTER2012370,WAXENEGGER2015138}. The geometry of the modeled PBs matched the design parameters of the fabricated structures. The dielectric function of gold was taken from Ref.~\cite{PhysRevB.6.4370} and the dielectric constant of the silicon nitride membrane was set equal to 4, which is a standard approximation in the considered spectral region.

{\it Results: Loss probability} We have fabricated two series of PBs, the first one with the fixed termination radius $r_\mathrm{fix}=70~\mathrm{nm}$ and the variable radius $r_\mathrm{var}$ varied from 70~nm to 20~nm in steps of 10~nm, and the second one with $r_\mathrm{fix}=50~\mathrm{nm}$ and $r_\mathrm{var}$ varied from 50~nm to 20~nm in steps of 10~nm. The length of all PBs was fixed at $L=300~\mathrm{nm}$. The actual dimensions might slightly differ from these design dimensions. For each combination of $r_\mathrm{fix}$ and $r_\mathrm{var}$, ten individual PBs were fabricated and three with the best correspondence to the designed shape were selected for further analysis.

\begin{figure}[ht!]
\includegraphics[clip,width=\columnwidth]{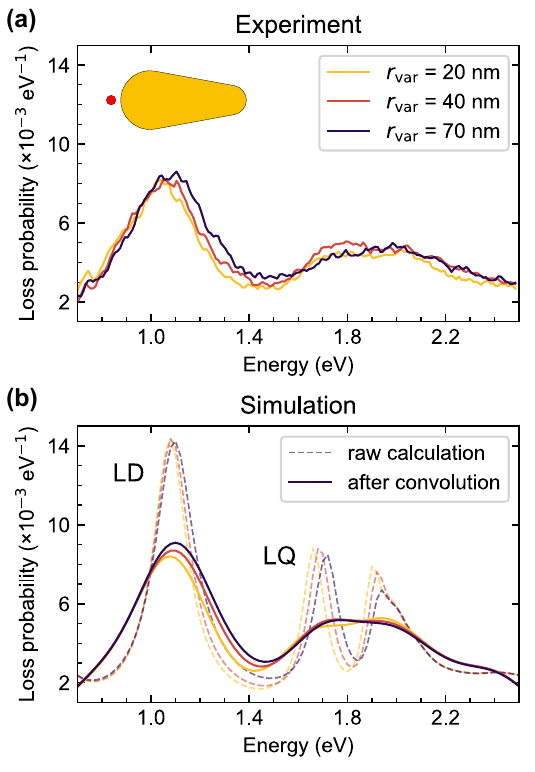}
\caption{\label{fig2}
Loss probability spectra for PBs with $r_\mathrm{fix}=70~\mathrm{nm}$ and the electron beam position indicated in the inset (20~nm from the PB boundary). (a) Experimental spectra for three values of $r_\mathrm{var}$. (b) Calculated spectra (solid lines) and calculated spectra convolved with a Gaussian (full width at half maximum of 0.27~eV) representing instrumental broadening.}
\end{figure}

\begin{figure}
\includegraphics[clip,width=\linewidth]{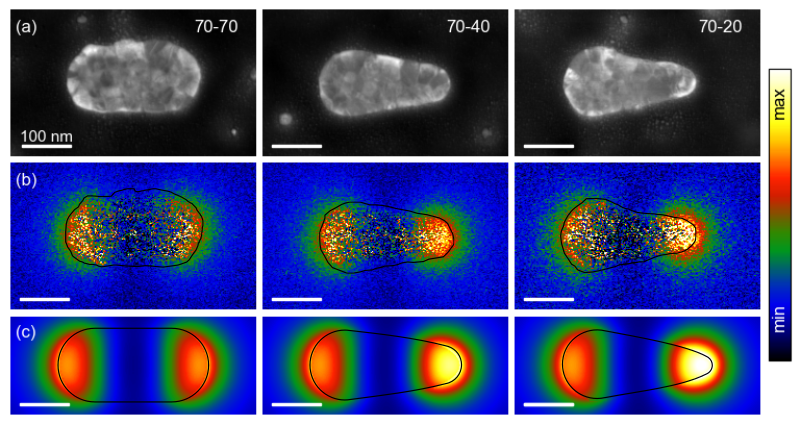}
\caption{\label{fig3}
Loss probability maps for the same PBs as in Fig.~\ref{fig2} at the energy of the longitudinal dipole LSP resonance (approximately 1.1~eV). (a) HAADF images of PBs. The numbers in the upper right corners represent $r_\mathrm{fix}$-$r_\mathrm{var}$ in nanometers. (b) Experimental maps averaged over the energy window with a width of 0.1~eV. (c) Calculated maps. Black contours in panels (b) and (c) represent the boundaries of PBs.}
\end{figure}

The loss probability is represented in the spectral domain in Fig.~\ref{fig2} and in the spatial domain in Fig.~\ref{fig3}. The experimental loss probability spectra [Fig.~\ref{fig2}(a)] exhibit a pronounced LSP resonance at the energy of about 1.1~eV identified as the longitudinal dipole (LD) resonance. This identification is based on the spatial distribution of the loss probability (Fig.~\ref{fig3} and calculated electric field (not shown), as explained in Ref.~\cite{krapek_independent}. The LD resonance is spectrally well separated from the higher-order resonances. By fitting Lorentzian profiles, we obtained principal LD resonance parameters: the resonance energy and the Q factor (defined as the resonance energy divided by the full width at the half maximum). \cc{U nasledujici vety zvazit kvantifikaci} These parameters are within the experimental error independent of the position of the electron beam, which proves that the loss probability is in this spectral region dominated by single LSP resonance. In the following, we will focus on the LD resonance, i.e., all quantities will be evaluated for the LD resonance or at its central energy.

Interestingly, the global parameters of the LD resonance (the resonance energy and the Q factor) are very similar for all PBs in both sets, with the energy in the interval 1.04\,--\,1.07~eV and the Q factors in the interval \cc{check the numbers}. In other words, changing $r_\mathrm{var}$ might influence the local response of PBs, but does not affect global LSP resonance parameters. This finding underlines the suitability of the PB sets to isolate the PLRE. 

In contrast to the global parameters of the LD resonance, the peak amplitude of the loss probability at the LD resonance energy varies with the position of the electron beam (as shown in Fig.~\ref{fig3}). The loss probability is closely related to the near electric field induced in PBs by the probe electron, as discussed in Refs.~\cite{PhysRevLett.100.106804,PhysRevLett.103.106801,C3CS60478K,krapek_independent}. The observation of the increased loss probability near the variable termination of the PB as the $r_\mathrm{var}$ is reduced [Fig.~\ref{fig3}(b), left to right, right side of the PBs] already demonstrates the validity of the PLRE. 

Theoretical loss probability [Fig.~\ref{fig2}(b), Fig.~\ref{fig3}(c)] reproduces the experimental loss probability [Fig.~\ref{fig2}(a), Fig.~\ref{fig3}(b)] \cc{upravit odkazy na obr. a panely podle finalni podoby} with a full qualitative and a good quantitative agreement. \cc{odkaz na sekci v supplementu} The experimental EELS therefore provides a validation of our theoretical simulations, which provide additional characteristics unavailable in the experiment, in particular, full vectorial electric field related to the LSP resonances.

\begin{figure}[ht!]
\includegraphics[clip,width=\linewidth]{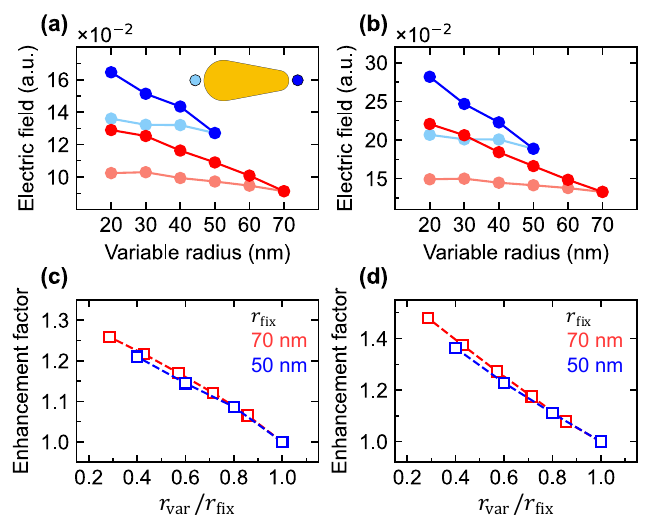}
\\
\caption{\label{fig4}
(a,b) Calculated test field $E_\mathrm{test}$ (dark blue/red) and reference field $E_\mathrm{ref}$ (light blue/red) as a function of the variable radius $r_\mathrm{var}$ for the fixed radius $r_\mathrm{fix}$ of 50~nm (blue) and 70~nm (red) at the distance from the PB $x_0$ of 10~nm (a) and 20~nm (b). The inset indicates the positions where the field is recorded.\cc{taken,evaluated?} (c,d) Enhancement factors $\mathit{EF}$ obtained for the fields from panels (a) and (b), respectively.}
\end{figure}

{\it Results: Electric field}
Next, we will discuss the electric field induced by the probe electron due to the excitation of the LD LSP resonance in the PB. We stress that the induced field does not include the excitation field of the probe electron; it is calculated as the total field from which the field of the electron in the medium without any PB is subtracted. We examine the test field $E_\mathrm{test}$ recorded at the variable-radius termination (where the PLRE effect varies) and the reference field $E_\mathrm{ref}$ recorded at the fixed-radius termination (where PLRE is constant; any variations in the reference field shall be attributed to the variations in additional effects contributing to the field magnitude, such as the evanescent confinement or the charge reservoir). \cc{zavorka je mozna zbytecne upovidana, zkontrolovat, zda to neopakujeme prilis} We evaluate the field at the long axis of the PB, midheight, and at the distance $x_0$ from the PB boundary, as indicated in the inset in Fig.~\ref{fig4}(a); the field value obtained right at the surface would suffer from the large numerical error.

In the calculations of the field, we set the electron beam so that it crosses the long axis of the PB 20 nm away from the {\it opposite} termination to that one where the field is recorded (i.e., at the variable-radius termination for $E_\mathrm{ref}$ recorded at the fixed-radius termination and vice versa). By the large distance between the probe electron and the recorded field, we suppress the numerical instabilities of the simulations.

The test and the reference electric field magnitudes are plotted in Fig.~\ref{fig4}(a,b) as functions of the variable radius $r_\mathrm{var}$ for two values of the distance $x_0$. We clearly observe that as the radius decreases (i.e., the curvature increases), the test field increases. The reference field varies only weakly with $r_\mathrm{var}$, but some variations are still present, indicating the influence of additional effects. \cc{tento pojem musime nekde zavest, at ho muzeme jednotne pouzivat} To remove them and isolate the pure effect of the local curvature, we have defined the enhancement factor as the ratio of the test and reference field, $\mathit{EF}=E_\mathrm{test}/E_\mathrm{ref}$, and plot it as a function of relative radius $r_\mathrm{var}/r_\mathrm{fix}$ in Fig.~\ref{fig4}(c,d). We observe an unequivocal increase of the enhancement factor for decreased relative radius, a clear manifestation of the PLRE.

At first glance, the values of the $\mathit{EF}$ seem to be rather low. To illustrate this impression, we will inspect the case of $r_\mathrm{fix}=70~\mathrm{nm}$, $r_\mathrm{var}=20~\mathrm{nm}$, and $x_0=20~\mathrm{nm}$, where we find the PLRE enhancement factor $\mathit{EF}=1.26$. \cc{zkontrolovat cislo} For the electrostatic lightning-rod effect (ELRE) in the geometry of Fig.~\ref{fig1}(a) we would obtain $\mathit{EF}_\mathrm{ELRE}=(r_\mathrm{fix}+x_0)/(r_\mathrm{var}+x_0)=2.25$. A natural interpretation of this observation would be that the PLRE is weak compared to ELRE. However, this is not correct. Instead, as we will show, the effective radii of curvature related to the spatial distribution of the induced charge are considerably smaller than the geometric radii. In the case considered above, we obtain
$r_\mathrm{fix}^\mathrm{eff}=16~\mathrm{nm}$ and $r_\mathrm{var}^\mathrm{eff}=6~\mathrm{nm}$, yielding the effective enhancement factor for the ELRE reading $\mathit{EF}_\mathrm{ELRE}^\mathrm{eff}=1.4$, comparable to the $\mathit{EF}=1.26$ for PLRE. 
\cc{zmeneno na rozsah fitu 10-150 nm}
\cc{nevim, zda je jasne, ze vezmeme hodnoty reff a dosadime do vztahu pro ELRE}

In doing so, we will formulate a phenomenological model for the induced field (restricting to the field at the long axis of the PB). We have tested several such models and we present here the minimum model, surprisingly simple, which reliably reproduces the real field. It consists of a single effective charge $Q_\mathrm{eff}$ induced inside the PB at the distance $r_\mathrm{eff}$ from the PB boundary (the so-called effective radius). \cc{efektivni polomer muze byt diskutovan jinde} We include the so-called dimensionality factor $d$ which describes how fast is the decay of the field into the dielectric, with the values of 2 and 1 corresponding to ideal point and linear charges, respectively. The equation for the electric field $E(x)$ in the dielectric \cc{tady taky musime sjednotit terminologii, nekde rikame air nebo vacuum} at the distance $x$ from the PB reads
\begin{equation}
E(x)=\frac{Q_\mathrm{eff}}{(r_\mathrm{eff}+x)^d}.
\end{equation}


\begin{figure}[ht!]
\includegraphics[clip,width=0.8\linewidth]{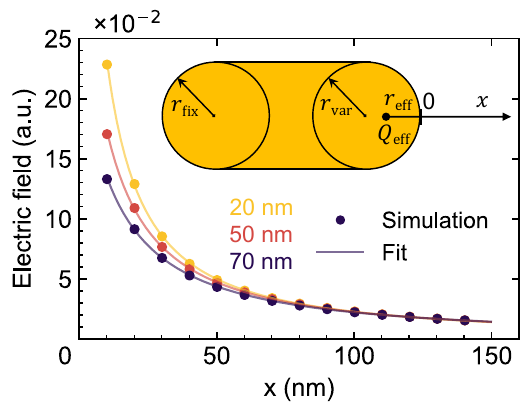}
\caption{\label{fig5}
The induced electric field near the variable-radius termination along the long axis of the PB (the line shown in the inset): the field obtained from electrodynamic simulations (symbols) and the fit by the phenomenological model (lines). The inset shows the parameters of the phenomenological model.}
\end{figure}


The profile of the induced electric field near the variable-radius termination (i.e., the test field) along the long axis of the PB is shown in Fig.~\ref{fig5}. The phenomenological model describes the field obtained from electrodynamic simulations very well. It is interesting to inspect the parameters of the model obtained from fitting the actual field (in the distance range from 10~nm to 150~nm). Importantly, the effective radius $r_\mathrm{eff}$ is very small, between 6~nm and 16~nm, and increases as $r_\mathrm{var}$ increases, ranging between $(0.2-0.3)\times r_\mathrm{var}$. In other words, the effective charge is positioned far from the center of the cylindric termination, near the edge of the PB. This is in striking contrast with the electrostatic case considered in Fig.~\ref{fig1}(b), where the charge exhibits a spherically symmetric distribution around the center of the sphere. In PLRE, the geometrical radius of the curvature of the PA significantly differs from the effective radius defined by the spatial distribution of the induced charge. Further, the dimensionality parameter $d$ is within the range of $1.2\pm 0.1$, i.e., corresponding rather to the linear effective charge. These facts explain rather quantitatively the strength of PLRE as shown in Fig.~\ref{fig4} and discussed above. \cc{prilis self-referenci, zkratit?}

The low effective radius has important consequences in the design of PAs. For example, when the field is to be enhanced 10~nm from a certain protrusion of the PA, reducing the radius of the protrusion from 50~nm to 20~nm (expected effective radius reduction from 15~nm to 6~nm for $r_\mathrm{eff}=0.3r$) is expected to increase the field by a fair factor of $25/16=1.6$, but further reduction to 10~nm (expected effective radius of 3~nm) brings expected reduction only by a factor of $16/13=1.2$ which might be not worth of the fabrication effort.



{\it Conclusion}
We have demonstrated the existence of the plasmonic lightning-rod effect, i.e., the formation of enhanced induced electric field related to a specific LSP resonance near highly-curved parts of the plasmonic antenna. By employing the sets of plasmonic beetroots allowing to compare the test and the reference field, we isolated the effect of the local curvature from other effects influencing the induced electric field, in particular the evanescent confinement and the charge reservoir. Quantitative analysis demonstrated that the plasmonic lightning-rod effect is comparably strong as the electrostatic lightning-rod effect, with the field exhibiting quasihyperbolic ($r^{-1}$) decay with the distance from the induced charge. Importantly, the charge is induced close to the boundary of plasmonic antennas, reducing their effective radius of curvature significantly in comparison to their geometrical radius. 


\begin{acknowledgments}
We acknowledge support from the Ministry of Education, Youth, and Sports of the Czech Republic, projects No.~CZ.02.01.01/00/22\_008/0004572 (QM4ST) and LM2023051 (CzechNanoLab).
\cc{specificky vyzkum? tady by to slo.}
\end{acknowledgments}

\bibliographystyle{apsrev4-2}
\bibliography{manuscript}

\end{document}